\documentclass[twocolumn,english,australian,pra]{revtex4-1}
\usepackage[T1]{fontenc}
\usepackage[latin9]{inputenc}
\usepackage{xcolor}
\usepackage{float}
\usepackage{amsmath}
\usepackage{amssymb}
\usepackage{graphicx}

\makeatletter

\newcommand*\LyXThinSpace{\,\hspace{0pt}}

\@ifundefined{definecolor}
 {\usepackage{color}}{}
\makeatother

\makeatother

\usepackage{babel}
\begin{document}

\title{Leggett-Garg tests of macro-realism for multi-particle systems including
two-well Bose-Einstein condensates}

\author{L. Rosales-Z\'arate$^{1}$, B. Opanchuk$^{1}$, Q. Y. He$^{2}$
and M. D. Reid$^{1}$}

\affiliation{$^{1}$Centre for Quantum and Optical Science, Swinburne University
of Technology, Melbourne 3122 Australia }

\affiliation{$^{\text{2}}$State Key Laboratory of Mesoscopic Physics, School
of Physics, Peking University, Beijing 100871 China}
\begin{abstract}
We construct quantifiable generalisations of Leggett-Garg tests for
macro/ mesoscopic realism and noninvasive measurability that apply
when not all outcomes of measurement can be identified as arising
from one of two macroscopically distinguishable states. We show how
quantum mechanics predicts a negation of the LG premises for proposals
involving ideal-negative-result, weak and quantum non-demolition measurements
on dynamical entangled systems, as might be realised with two-well
Bose-Einstein condensates, path-entangled NOON states and atom interferometers.
\end{abstract}
\maketitle
Schrodinger raised the apparent inconsistency between macroscopic
realism and quantum macroscopic superposition states \cite{Schrodinger-1}.
Leggett and Garg (LG) suggested to test macroscopic realism against
quantum mechanics in an objective sense by comparing the predictions
of quantum mechanics with those based on two very general classical
premises  \cite{LG}. The first premise is\emph{ macroscopic realism}
(MR), that a system which has two macroscopically distinguishable
states available to it is at any time in one or other of the states.
The second premise is \emph{noninvasive measurability} (NIM), that
for such a system it is possible to determine which state the system
is in, without interfering with the subsequent evolution of that system.

Leggett and Garg  showed how the two premises constrain the dynamics
of a two-state system. Considering three successive times $t_{3}>t_{2}>t_{1}$,
the variable $S_{i}$ denotes which of the two states the system is
in at time $t_{i}$, the respective states being denoted by $S_{i}=+1$
or $-1$. The LG premises impl\textcolor{black}{y the LG inequality
\cite{LG,jordan_kickedqndlg2}}
\begin{eqnarray}
LG=\langle S_{1}S_{2}\rangle+\langle S_{2}S_{3}\rangle-\langle S_{1}S_{3}\rangle & \leq & 1.\label{eq:lg2}
\end{eqnarray}
and also the ``disturbance'' or ``no signalling in time (NST)''
inequality $d_{\sigma}=\langle S_{3}|\hat{M}_{2},\sigma\rangle-\langle S_{3}|\sigma\rangle=0$
\cite{NSTmunro,nst}. Here $\langle S_{3}|\hat{M}_{2},\sigma\rangle$
(and $\langle S_{3}|\sigma\rangle$) is the expectation value of $S_{3}$
given that a measurement $\hat{M}_{2}$ is performed (or not performed)
at time $t_{2}$, conditional on the system being prepared in a state
denoted $\sigma$ at time $t_{1}$. These inequalities can be violated
for quantum systems \cite{Emary_Review_LG,LG,lgexpphotonweak,idealnegexp,guo-qcoh,weak_backaction_resolution,Munro,nst,NSTmunro,jordan_kickedqndlg2,Mitchell,massiveosci}.
\textcolor{blue}{}The work of LG represented an advance, since it
extended beyond the quantum framework to show how\emph{ }the macroscopic
superposition defies classical macroscopic reality.

The LG approach raised new ideas about how to test quantum mechanics
even at the microscopic level  \cite{idealnegexp,lgexpphotonweak,Emary_Review_LG,guo-qcoh,weak_backaction_resolution}.
Failure of the inequalities implies no classical trajectory exists
between successive measurements: either the system cannot be viewed
as being in a definite state independent of observation, or there
cannot be a way to determine that state, without interference by the
measurement. Noninvasive measurability would seem ``vexing'' to
justify, however, because of the plausibility of the measurement
process disturbing the system. LG countered this problem by proposing
an \emph{ideal negative result measurement} (INR): the argument is
conditional on the first postulate being true e.g. \emph{if} a photon
does travel through one slit\emph{ or} the other, a null detection
beyond one slit is justified to be noninvasive \cite{LG,idealnegexp}.
A second approach is to perform \emph{weak measurements} \cite{weak,weakLGbellexp_review}
that enable calculation of the moment $\langle S_{2}S_{3}\rangle$
in a limit where there is a vanishing disturbance to the system \cite{weakLGbellexp_review,Emary_Review_LG,lgexpphotonweak,weak_backaction_resolution}.
To date, experimental investigations involving INR or weak measurements
have focused mostly on microscopic systems e.g. a single photon. An
exception is a recent experiment which gives evidence for violation
of MR using a simpler form of LG inequality that quantifies the invasiveness
of ``clumsy'' measurements, and is applied to superconducting flux
qubits \cite{NSTmunro}. There have also been recent proposals for
LG tests involving macroscopic mechanical oscillators \cite{massiveosci}
and for macroscopic states of cold atoms, using  \emph{quantum non-demolition}
(QND) measurements \cite{Mitchell}\textcolor{blue}{.}\textcolor{black}{}\textcolor{red}{{}
}

An illuminating LG test would be for a mesoscopic massive system in
a quantum superposition of being at two different locations \cite{penrosediosi}.
An example of such a superposition is the path-entangled NOON state,
written as $|\psi\rangle=\frac{1}{\sqrt{2}}\{|N\rangle_{a}|0\rangle_{b}+|0\rangle_{a}|N\rangle_{b}\}$
where $|N\rangle_{a/b}$ is the $N$-particle state for a mode $a$
($b$) \cite{opticalNOON}. In this case the ideal negative result
measurement can be applied, and justified as noninvasive by the assumption
of Bell's locality \cite{Bell}. A method is then given to (potentially)
negate that the system must be located either ``here'' or ``there'',
or else to conclude there is a significant disturbance to a massive
system due to a measurement performed at a different location.

In this paper, we show how such tests may be possible on a mesoscopic
scale. As one example, we show that LG violations are predicted for
Bose-Einstein condensates (BEC) trapped in two separated potential
wells of an optical latttice. Here dynamical oscillation of large
groups of atoms to form NOON macroscopic superposition states is predicted
at high nonlinearities \cite{carr,carrmacrosup,two-wellent,carr-ent,esteve}. 

A key problem however for an actual experimental realisation is the
fragility of the macroscopic superpositions. To address this problem,
we derive \emph{modified LG inequalities} that can be used to test
LG premises for superpositions  that deviate from the ideal NOON
superposition by allowing mode population differences not equal to
$-N$ or $N$. The ideal negative result measurement is difficult
to apply where there are residual atoms in both modes, and we thus
develop weak and QND measurement strategies for testing LG premises
and demonstrating mesoscopic quantum coherence in this case. Tests
of LG realism are also possible for NOON states incident on an interferometer.
Finally, we propose a simple LG test for matter waves passing through
an atom interferometer could demonstrate the no-classical trajectories
result for atoms.

\emph{Idealised dynamical two-state oscillation:} The Hamiltonian
$H_{I}$ for an $N$-atom condensate constrained to a double well
potential reveals a regime of macroscopic two-state dynamics. The
two-well system has been reliably modelled by the Josephson two-mode
Hamiltonian \cite{josHam,carr,esteve,grossnonlinearint}:
\begin{equation}
H_{I}=2\kappa\hat{J}_{z}+g\hat{J}_{z}^{2}\label{eq:Hamiltonian-1-1-1-1}
\end{equation}
Here $\hat{J}{}_{z}=(a^{\dagger}a-b^{\dagger}b)/2$ , $\hat{J}{}_{x}=(a^{\dagger}b+b^{\dagger}a)/2$,
$\hat{J}{}_{y}=(a^{\dagger}b-b^{\dagger}a)/2i$ are the Schwinger
spin operators defined in terms of \textcolor{black}{the boson operators
$\hat{a}^{\dagger},\hat{a}$ and $\hat{b}^{\dagger},\hat{b}$, for
the modes describing particles in each of the wells, labelled $a$
and $b$ respectively.}\textcolor{red}{{} }\textcolor{black}{The $\kappa$
models interwell hopping and $g$ the nonlinear self-interaction due
to the medium. In a regime of high interaction }strength ($Ng/\kappa\gg1$\textcolor{black}{),
a regime exists where if the system is initially prepared with all
$N$ atoms in one well, a two-state oscillation can take place with
period $T_{N}$ (Fig. 1)} \cite{carrmacrosup,carr}. In one state,
$|N\rangle_{a}|0\rangle_{b}$\emph{, all} $N$ atoms are in the well
$a$ ($S_{i}=1$), and in the second state, $|0\rangle_{a}|N\rangle_{b}$\emph{,
all }atoms are in the well $b$ ($S_{i}=-1$) \cite{carr}. If the
system is prepared in $|N\rangle_{a}|0\rangle_{b}$\emph{,} then at
a later time $t'$, the state vector is \textcolor{red}{ }\textcolor{black}{(apart
from phase factors)}\textcolor{red}{{} }
\begin{eqnarray}
|\psi(t)\rangle & = & \cos(\tau)|N\rangle|0\rangle+\sin(\tau)|0\rangle|N\rangle\label{eq:twostate-1}
\end{eqnarray}
where $\tau=E_{\Delta}t'/\hbar$ and $E_{\Delta}$ is the energy splitting
of the energy eigenstates $|N\rangle|0\rangle\pm|0\rangle|N\rangle$
 under $H_{I}$. 

The quantum solution (\ref{eq:twostate-1}) predicts a violation of
the LG inequality. The two-time correlation is $\langle S_{i}S_{j}\rangle=\cos\left[2(t_{j}-t_{i})\right]$
and is independent of the initial state, whether $|N\rangle|0\rangle$
or $|0\rangle|N\rangle$. Choosing $t_{1}=0$, $t_{2}=\pi/6$, $t_{3}=\pi/3$
(or $t_{3}=5\pi/12$), it is well-known that for this two-time correlation
the quantum prediction is $LG=1.5$ ($1.37$) which gives a violation
of (\ref{eq:lg2}) \cite{LG}.

\textcolor{black}{The tunnelling times in the highly nonlinear regime
however are impractically high for proposals based on Rb atoms \cite{tuntime,oberoscexp}.
The fragility of the macroscopic superposition state will make any
such experiment unfeasible \cite{deco}. Noting however that the modes
$a$, $b$ of $H_{I}$ may also describe occupation of two atomic
hyperfine levels, $\kappa$ being the Rabi frequency as in the experiments
of \cite{grossnonlinearint}, the NOON oscillation may well be achievable
for other physical realisations of $H_{I}$. }Alternatively, for
more practical oscillation times one can use a different initial state
$|N-n_{L}\rangle|n_{L}\rangle$, $0<n_{L}<N$, where there are atoms
in both wells, or else a tilted well \cite{carr}. Here, we denote
the sign of the spin $J_{z}$ at time $t_{i}$ by $S_{i}$ ($S_{i}=1$
if $J_{z}\geq0$; $S_{i}=-1$ if $J_{z}<0$). The dynamical solutions
presented in Fig. 2  reveal a mesoscopic two-state oscillations over
reduced time scales, mimicking the experimentally observations of
Albiez et al \cite{oberoscexp} for $N=1000$ atoms where oscillations
were observed over milliseconds. 
\begin{figure}[H]
\begin{centering}
\par\end{centering}
\begin{centering}
\includegraphics{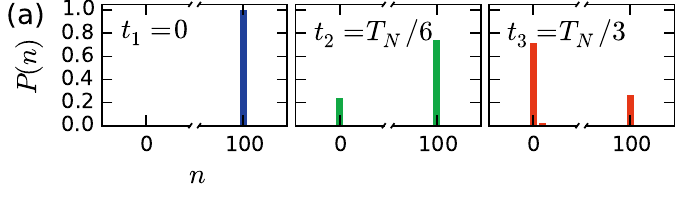}\\
\includegraphics{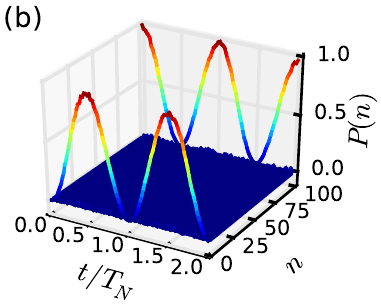}\ \,\ \includegraphics{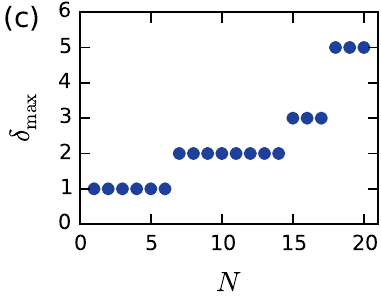}
\par\end{centering}
\centering{}\caption{\textcolor{black}{\emph{Oscillating NOON two-state dynamics}}\textcolor{black}{:
(a) Probabilities of $n$ atoms in mode $a$ at times $0,\ T_{N}/6,\ T_{N}/3$}
and (b)\textcolor{red}{{} }the two-state oscillation for $N=100$, $g=1$.
(c) Plot gives an upper bound on the  backaction $\delta$ due
to the INR measurement that can be tolerated for an LG violation.
\textcolor{red}{}\textcolor{red}{}\textcolor{blue}{\label{fig:ideal}}\textcolor{red}{}}
\end{figure}

Our objective is to provide practical strategies for testing the
LG inequality in such multiparticle experiments. Two questions to
be addressed are how to perform (or access the results of) the NIM
(assuming it exists), and how to handle the case where the values
of $S_{i}$ may not always correspond to macroscopically distinct
outcomes.

To address the first question: As explained in the literature \cite{LG},
\textcolor{black}{$\langle S_{1}S_{2}\rangle$ and $\langle S_{1}S_{3}\rangle$
can be inferred using deterministic state preparation and projective
measurements at $t_{2}$ and $t_{3}$. To measure $\langle S_{1}S_{3}\rangle$
no intervening measurement is made at $t_{2}$ based on the assumption
that the NIM at $t_{2}$ will not affect the subsequent statistics.}
\textcolor{black}{For $\langle S_{2}S_{3}\rangle$,  }the evaluation
of $S_{2}$ is difficult, since with any practical measurement it
could be argued that a measurement $M$ made at $t_{2}$ is not the
NIM, and does indeed influence the subsequent dynamics. Three methods
have been used to counter this objection: INR measurements; weak
measurements; and quantifiable QND measurements. We next propose
LG tests for each case.

(1) \emph{Ideal negative result measurement (INR):} A particularly
strong test is possible for experiments involving a NOON superposition
(3) where the two modes correspond at time $t_{2}$ to spatially separated
locations. In this case, the INR strategy similar to that outlined
by LG can be applied. A measurement apparatus at time $t_{2}$ couples
locally to only one mode $a$, enabling measurement of the particle
number $n_{a}$. Either $n_{a}=0$ or $n_{a}=N$. Based on the first
LG premise, if one obtains the negative result $n_{a}=0$, it is assumed
that there were prior to the measurement \emph{no} atoms in the mode
$a$. Hence the measurement that gives a negative result is justified
to be noninvasive (since $\langle S_{2}S_{3}\rangle$ can be evaluated
using only negative result outcomes \cite{LG}). For such an experiment,
to assume noninvasive measurability there is implicit the assumption
of locality: that there is no change to mode $b$ because of the measurement
at $a$ (otherwise a change to the subsequent dynamics could be expected).
\begin{figure}[t]
\includegraphics{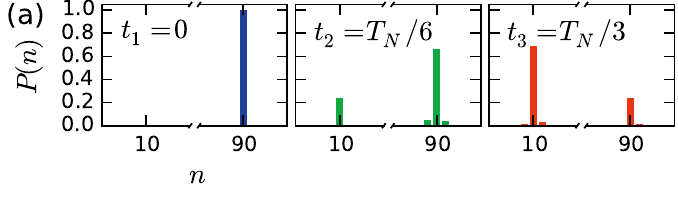}\\
\includegraphics{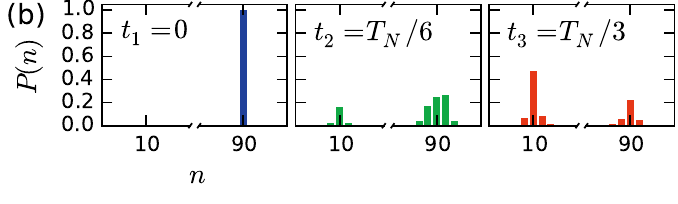}

\caption{\emph{Mesoscopic two-state oscillations:}\textcolor{red}{{} }Probability
of $n$ atoms in mode $a$ at $t_{2}$ and $t_{3}$. $N=100$. The
initial state\textbf{ }is $n_{L}=10$ particles in the right well.
Here (a) $g=3$ and (b) $g=1$. (a) $LG=1.49$ using the weak measurement
of $\hat{J}_{z}$ at $t_{2}$ \textcolor{red}{}\textcolor{red}{}and
(b)\textcolor{purple}{{} }$LG=1.43$ using the minimal QND measurement
of $S_{2}$ at $t_{2}$.\textcolor{red}{}\textcolor{black}{{} }\textcolor{black}{\label{fig:nonideal N=00003D10}}\textcolor{red}{}}
\end{figure}

\emph{Quantification of the NIM premise:} We can introduce a quantification
of the second LG premise: We suppose that the measurement at mode
$a$ ($b$)\emph{ can }induce a back-action effect on the macroscopic
state of the other mode, so that there may be a change of the state
of mode $b$ ($a$) of up to $\delta$ particles, where $\delta\leq N$.
The change $\delta$ may be microscopic, not great enough to switch
the system between states $|0\rangle|N\rangle$ and $|N\rangle|0\rangle$,
but can alter the subsequent dynamics. The change to the dynamics
is finite however, and can be established within quantum mechanics,
to give a range of prediction for $\langle S_{2}S_{3}\rangle$. We
have carried out this calculation, and plot the effect of $\delta$
for various $N$ in Fig. 1c, noting that a relatively small backaction
$\delta$ to the quantum state of one mode due to measurement on the
other will destroy violations of the LG inequality even for large
$N$.

(2) \emph{Weak and minimally invasive QND measurements:} A second
strategy is to construct a measurement that can be shown to give a
negligible disturbance to the system being measured. We consider
a QND measurement described by the Hamiltonian $H_{Q}=\hbar G\hat{J}_{z}\hat{n}_{c}$.
The $H_{Q}$ models QND measurements of the atomic spin $\hat{J}_{z}$
based on an ac Stark shift \cite{qndPCI}. An optical ``meter''
field is prepared in a coherent state $|\gamma\rangle$ and coupled
to the system for a time $\tau_{0}$. The meter field is a single
mode with boson operator $\hat{c}$ and number operator $\hat{n}_{c}=\hat{c}^{\dagger}\hat{c}$.
Writing the state of the system at time $t_{2}$ as $\sum_{m=0}^{N}d_{m}|m\rangle_{a}|N-m\rangle_{b}$
($d_{m}$ are probability amplitudes), the output state immediately
\emph{after} measurement is (we set $\tau_{0}=\pi/2NG$)\textcolor{blue}{}
\begin{eqnarray}
|\psi\rangle & = & \sum_{m=0}^{N}d_{m}|m\rangle_{a}|N-m\rangle_{b}|\gamma e^{i\pi(N-2m)/2N}\rangle_{c}\label{eq:weakstate}
\end{eqnarray}
Homodyne detection enables measurement of the meter quadrature phase
amplitude $\hat{p}=(\hat{c}-\hat{c}^{\dagger})/i$. For $\gamma$
large, the measurement is ``strong'', or projective, and the different
values of $\hat{J}_{z}$ (and hence $S_{2}$) are precisely measurable
as distinct regions of outcomes for $\hat{p}$. Ideally, a \emph{minimal
QND} measurement of $S_{2}$ is devised, that does not discriminate
the different values of $\hat{p}$ apart from the sign, and leaves
all states (\ref{eq:weakstate}) with definite sign $S_{2}$ unchanged.

First, we discuss the\emph{ weak} measurement limit, attained as $\gamma\rightarrow0$.
In this limit, we see from (\ref{eq:weakstate}) that the state of
the system is \emph{minimally disturbed} by the measurement. The cost
is no clear resolution of the value $S_{2}$ for any single measurement.
Yet, using (\ref{eq:weakstate}), we show in the Supplemental Materials
that if the system at time $t_{2}$ is in a NOON state  $d_{0}|N-n_{L}\rangle_{a}|n_{L}\rangle_{b}+d_{N}|n_{L}\rangle|N-n_{L}\rangle$,
then\textcolor{purple}{{} }\textcolor{red}{}
\begin{eqnarray}
\langle S_{2}S_{3}\rangle & = & -\frac{1}{2\gamma}\langle pS_{3}\rangle\label{eq:weak}
\end{eqnarray}
\textcolor{black}{where $\langle S_{2}S_{3}\rangle$ }\textcolor{blue}{}\textcolor{black}{is
the value obtained by the projective measurement. }\textcolor{black}{}Thus,
for arbitrarily small $\gamma$, the value $\langle S_{2}S_{3}\rangle$
can be obtained by averaging over many trials.  The weak measurement
strategy enables a convincing test of the LG premises, since one can
experimentally demonstrate the noninvasiveness of the weak measurement,
by showing the invariance of $\langle S_{1}S_{3}\rangle$ as $\gamma\rightarrow0$
when the measurement is performed at $t_{2}$. 

The weak measurement relation (\ref{eq:weak}) does not hold for all
input states. However, the \emph{minimal} (``non-clumsy'') QND measurement
of $S$ gives a strategy for LG tests, based on extra assumptions.
For systems such as in Fig. 2, the state at time $t_{2}$ is a superposition
of states $|\psi_{+}\rangle$ and $|\psi_{-}\rangle$ that give,
respectively, outcomes $S=\pm1$.  The first LG premise is that the
system is \emph{either} in a state of positive $S$ \emph{or} in a
state with negative $S$. The minimal QND strategy requires a second
set of measurements, in order to experimentally establish that states
with definite value of $S$ are unchanged by the QND measurement \textcolor{black}{\cite{Mitchell,clumsy-meas,NSTmunro}}.
The noninvasiveness of the measurement is then justified by the first
LG premise. \textcolor{black}{We note that in cases where the measurement
is not ideal (``clumsy''), the amount of disturbance can be measured
and accounted for in a modified inequality as discussed in the Refs.
\cite{Mitchell,clumsy-meas,NSTmunro}.} Strictly speaking, the QND
approach is limited to testing a modified LG assumption  that the
system is always in a \emph{quantum} state with definite $S$ at the
time $t_{2}$. This is because it is difficult to prove that all hidden
variable states with definite outcome of $S$ are not changed by the
QND measurement. Regardless, the approach rigorously demonstrates
the \emph{quantum coherence} between the states $|\psi_{+}\rangle$
and $|\psi_{-}\rangle$. Fig. 2 shows LG violations using weak and
QND measurements.
\begin{figure}[t]

\includegraphics[width=0.5\columnwidth]{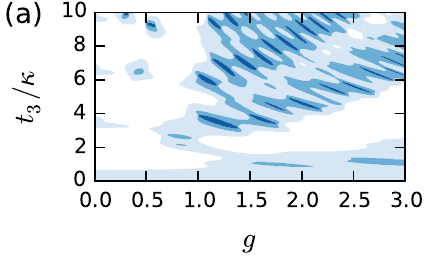}\includegraphics{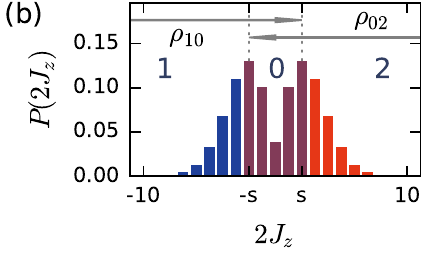}

\caption{\textcolor{black}{\emph{Violation of $s$-scopic LG inequalities}}\emph{
using nonlinear interferometers:}\textcolor{black}{{} }\textcolor{red}{}\textcolor{black}{\label{fig:nonideal N=00003D10-1}}\textcolor{red}{}\textcolor{black}{}\textbf{\textcolor{red}{}}\textcolor{black}{(a)
The NOON state (3) is created at time $t_{2}$  ($N=5$, $\tau=\pi/6$)
and evolves to time $t_{3}$ according to $H_{I}$ with nonlinearity
$g$.}\textcolor{purple}{{} }\textcolor{black}{}\textcolor{red}{}\textcolor{black}{Contours
show regimes for violation of the $s$-scopic inequality}\textcolor{purple}{{}
}\textcolor{black}{(\ref{eq:LGgen-1}) where (dark to light) $s=4,2,0$.
}\textcolor{red}{}\textbf{\textcolor{red}{}}\textcolor{black}{(b)}
Schematic of the probability distribution for results $2J_{z}$
\textcolor{black}{at $t_{3}$} depicting three regions $1$, $0$,
and $2$. }
\end{figure}

\emph{The $s$-scopic LG inequalities:} We now address how to test
macroscopic realism where the system deviates from the ideal of two
macroscopically distinguishable states.  This occurs when there is
a nonzero probability for $J_{z}$ different to $\pm N/2$ as in Figure
3b. Adapting the approach put forward by LG and Refs. \cite{cavalreiduncer2006},
we define three regions of $J_{z}$: region ``$1$'', $J_{z}<-s/2$;
region ``$0$'', $-s/2\leq J_{z}\leq s/2$; and region ``$2$'',
$J_{z}>s/2$. \textcolor{blue}{} 

For arbitrary $s$, the MR assumption is accordingly renamed, to \emph{s-scopic
realism }(sR). In the generalised case, the meaning of sR is
that the system is in a probabilistic mixture of two \emph{overlapping}
states: the first that gives outcomes in regions ``$1$'' or ``$0$''
(denoted by hidden variable $\tilde{S}=-1$); the second that gives
outcomes in regions ``$0$'' or ``$2$'' (denoted by $\tilde{S}=1$).
The second LG premise is generalised to \emph{$s$-scopic noninvasive
measurability} which asserts that such a measurement can be made
at time $t_{2}$ without changing the result $J_{z}$ at time $t_{3}$
by an amount $s$ or more.

\textcolor{black}{The $s$-scopic LG premises imply a quantifiable
inequality. This is because  any effects due to the overlapping region
are limited by the finite probability of observing a result there.
}Defining the measurable marginal probabilities of obtaining a result
in region $j\in\left\{ 0,1,2\right\} $ at the time $t_{k}$ by $P_{j}^{(k)}$,
the $s$-scopic premises are violated if \cite{sm}\textcolor{red}{}
\begin{equation}
LG_{s}=P_{2}^{(2)}-P_{1}^{(2)}+\langle S_{2}S_{3}\rangle-(P_{2}^{(3)}-P_{1}^{(3)})-2P_{0|M}^{(3)}-P_{0}^{(3)}>1\label{eq:LGgen-1}
\end{equation}
\textcolor{black}{where we have used that the system is prepared initially
in region 2 and }here we restrict to scenarios satisfying $P_{0}^{(2)}=0$.\textcolor{black}{{}
}The $\langle S_{2}S_{3}\rangle$ is to be measured using a noninvasive
measurement at $t_{2}$. \textcolor{red}{}A similar modification
is given for the disturbance inequality: The $s$R premises are violated
if $d_{\sigma,s}=|P_{2|M}^{(3)}-P_{1|M}^{(3)}-(P_{2}^{(3)}-P_{1}^{(3)})|-(P_{0}^{(3)}+P_{0|M}^{(3)})>0$
where $P_{j|M}^{(3)}$ ($P_{j}^{(3)}$) is the probability with (without)
the measurement $M$ performed at $t_{2}$.
\begin{figure}[t]

\includegraphics[width=0.75\columnwidth]{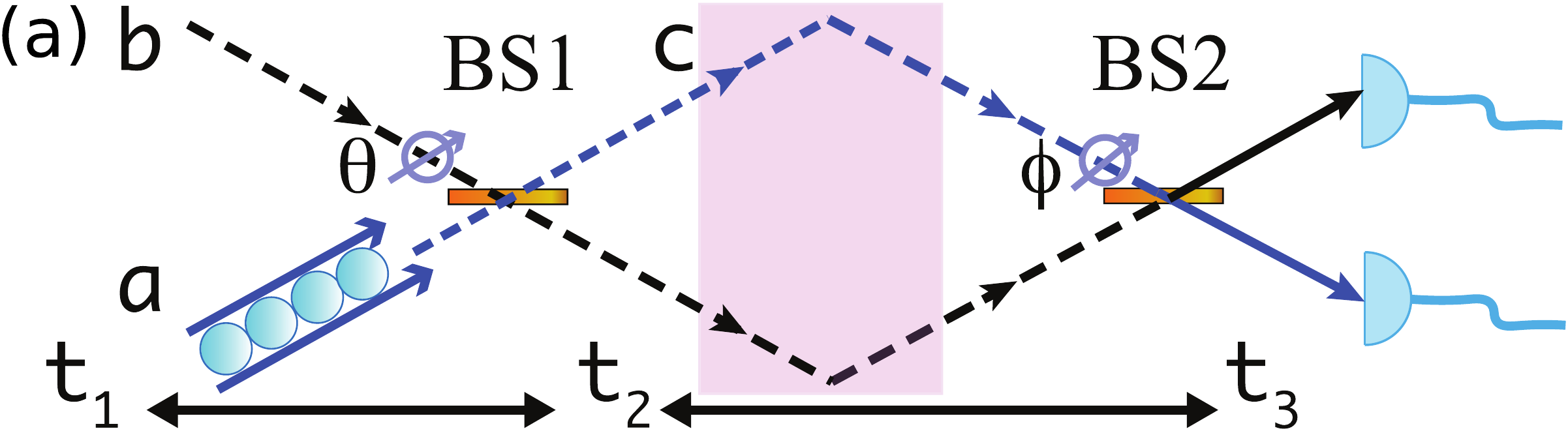}

\medskip{}

\textcolor{blue}{\includegraphics[width=0.5\columnwidth]{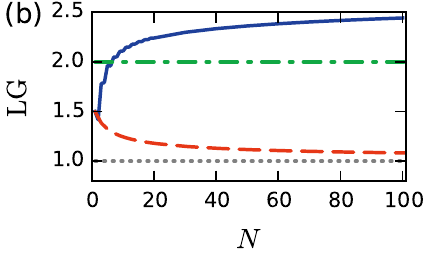}}\includegraphics[width=0.5\columnwidth]{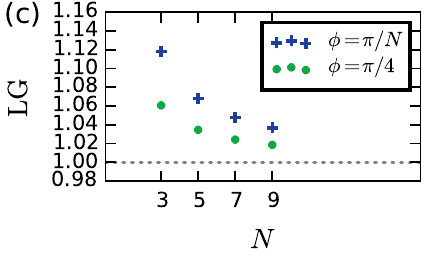}\\

\caption{\emph{Multiparticle linear interferometers:}\textcolor{black}{{} (a)}\textcolor{red}{{}
}\textcolor{black}{$N$ bosons pass through an interferometer. A
QND measurement $\hat{M}$ (purple shading) is made on the state created
at $t_{2}$.  (b) }(blue solid curve) Violation of LG inequality
for optimal rotation angles $\theta$, $\phi$ where $\hat{M}$\textcolor{black}{{}
measures the number of particles in arm $c$, showing ``no classical
trajectories''}\textcolor{red}{{} }\textcolor{black}{for individual
atoms.}\textcolor{red}{{} }\textcolor{black}{(b) (dashed curves)
LG violations when $\hat{M}$ is }\textcolor{red}{}a minimal (non-clumsy)
measurement of $S_{2}$. \textcolor{black}{Green dotted-dashed curve
shows the disturbance $d_{\sigma}=2$ (for all $\Delta<N$) where
mesoscopic} superposition states $|\psi_{\Delta}\rangle$ are created
at $t_{2}$\textcolor{black}{. Red dashed curve shows LG value for
odd $N$ where $\Delta=0$. For $\Delta=N-1$, the NOON state (3)
with $\tau=\theta$ }is created at $t_{2}$. $\hat{M}$ can then
be realised as an INR or weak measurement.\textbf{\textcolor{red}{{}
}}\textcolor{red}{}\textcolor{black}{}Fig (c) shows violations
in that case using a phase shift $\phi$ and a 50/50 $BS2$ (with
optimal $\theta$). \textcolor{red}{}\textcolor{black}{}\textcolor{red}{}\textcolor{black}{\label{fig:nonideal N=00003D10-1-2}}\textcolor{red}{}\textcolor{black}{}\textbf{\textcolor{red}{}}\textcolor{red}{}}
\end{figure}

\emph{Nonlinear and linear interferometers: }Figure 3a shows predictions
for $s$-scopic violations. A NOON state (3) is created at $t_{2}$
and a weak measurement or INR performed. The NOON state might be
created via the nonlinear $H_{I}$ or by the conditional methods that
have been \textcolor{black}{}applied to photonic states \cite{ConditionalNOON}.
Subsequently, the system evolves according to the nonlinear interaction
$H_{I}$ and a measurement is made of $J_{z}$ at $t_{3}$. For realistic
timescales, there is a spread of $J_{z}$ at the times $t_{3}$ (Fig.
3b). These regimes are realisable for finite $g$ and $N\sim100$
in BEC nonlinear interferometers \cite{grossnonlinearint}.\textcolor{red}{} 

LG tests with mesoscopic superposition/ NOON states are also possible
\emph{without nonlinearity} at $t>t_{2}$, if, after the weak/ INR
measurement at $t_{2}$, the two modes are combined across a variable-angle
beam splitter (or beam splitter with phase shift $\phi$) and $J_{z}$
of the outputs measured (Fig. 4).\textcolor{black}{{}  }\textcolor{red}{}\textcolor{black}{The
macroscopic Hong-Ou-Mandel technique conditions on $|J_{z}|>\Delta/2$
($\Delta<N$) to create at $t_{2}$ an $N$-atom mesoscopic superposition
state $|\psi_{\Delta}\rangle=|\psi_{+}\rangle+|\psi_{-}\rangle$ where
states $|\psi_{\pm}\rangle$ are distinct by more than $\Delta$ particles
in each arm of the interferometer \cite{ConditionalNOON}. Violations
of the disturbance and LG inequalities are plotted in Figs. 4b and
c. Results indicate small violations for $s\sim2$ over a range of
$N$ and $\Delta$ \cite{sm}.}

\emph{No-classical trajectories for atom interferometers:} Finally,
we propose a simple test to falsify classical trajectories in the
multi-particle case for simple interferometers. At $t_{1}$, $N$
particles pass through a polariser beam splitter (or equivalent) (BS1)
rotated at angle $\theta$ (Fig. 4). The number difference of the
outputs if measured indicate the value of $J_{\theta}$ (and $S_{2}$)
at $t_{2}$. The particles are then incident on a second beam splitter
$BS2$ at angle $\phi$ whose output number difference gives $S_{3}$
at $t_{3}$. \textcolor{black}{We invoke the premise, that the system
is always in a state of definite $J_{\theta}$ prior to measurement
at $t_{2}$. This is based on the hypothesis that }\textcolor{black}{\emph{each}}\textcolor{black}{{}
atom goes one way }\textcolor{black}{\emph{or}}\textcolor{black}{{}
the other, through the paths of the interferometer. }\textcolor{black}{A
second premise is also invoked, that a measurement could be performed
of $J_{\theta}$ at $t_{2}$ that does not disturb the subsequent
evolution.} \textcolor{black}{The} second premise is justified by
the first, and can be supported by experiments that create a spin
eigenstate, and then demonstrate the complete invariance of the state
after the QND number measurement.\textcolor{black}{} If the premises
are valid, the LG inequalities (\ref{eq:lg2}) will hold, but by contrast
are predicted violated by quantum mechanics (Fig. 4b (blue solid curve)).
\textcolor{black}{While not the macroscopic test LG envisaged, this
gives an avenue for workable tests of the ``classical trajectories''
hypothesis  that could be applied to atoms \cite{atomint}. }

\end{document}